\documentclass[preprint, 12pt]{elsarticle}

\usepackage{xcolor}
\usepackage{hyperref}
\usepackage{graphicx, color}
\usepackage{textcomp}
\usepackage{algorithm}
\usepackage{algpseudocode}
\usepackage{url}
\usepackage{amsthm}
\usepackage{subcaption}
\usepackage{booktabs}
\newtheorem{theorem}{Theorem}
\usepackage{tcolorbox}
\usepackage{mathtools}
\usepackage{amssymb}
\usepackage{multicol}
\usepackage{fix-cm}

\def\BibTeX{{\rm B\kern-.05em{\sc i\kern-.025em b}\kern-.08em
    T\kern-.1667em\lower.7ex\hbox{E}\kern-.125emX}}
\AtBeginDocument{\definecolor{tmlcncolor}{cmyk}{0.93,0.59,0.15,0.02}\definecolor{NavyBlue}{RGB}{0,86,125}}

\algnewcommand\algorithmicassert{\texttt{assert}}
\algnewcommand\algdelete{\texttt{delete}}
\algnewcommand\Assert[1]{\State \algorithmicassert #1}
\algnewcommand\Delete[1]{\State \algdelete #1}

\theoremstyle{definition}
\newtheorem{definition}{Definition}[section]

\begin{document}
\begin{frontmatter}
\author[udc]{David Soler\texorpdfstring{\corref{cor1}}{*}}
\ead{david.soler@udc.es}
\cortext[cor1]{Corresponding author}

\title{Federated Anonymous Blocklisting across Service Providers and its Application to Group Messaging}

\author[udc]{Carlos Dafonte}
\author[uvigo]{Manuel Fernández-Veiga}
\author[uvigo]{Ana Fernández Vilas}
\author[udc]{Francisco J. Nóvoa}

\begin{abstract}
Instant messaging has become one of the most used methods of communication online, which has attracted significant attention to its underlying cryptographic protocols and security guarantees. Techniques to increase privacy such as End-to-End Encryption and pseudonyms have been introduced. However, online spaces such as messaging groups still require moderation to prevent misbehaving users from participating in them, particularly in anonymous contexts.. In Anonymous Blocklisting (AB) schemes, users must prove during authentication that none of their previous pseudonyms has been blocked, preventing misbehaving users from creating new pseudonyms. In this work we propose an alternative \textit{Federated Anonymous Blocklisting} (FAB) in which the centralised Service Provider is replaced by small distributed Realms, each with its own blocklist. Realms can establish trust relationships between each other, such that when users authenticate to a realm, they must prove that they are not blocked in any of its trusted realms. We provide an implementation of our proposed scheme; unlike existing AB constructions, the performance of ours does not depend on the current size of the blocklist nor requires processing new additions to the blocklist. We also demonstrate its applicability to real-world messaging groups by integrating our FAB scheme into the Messaging Layer Security protocol.
\end{abstract}


\end{frontmatter}

\section{Introduction}


In recent years, instant messaging has become the main method of communication over the internet~\cite{messaging-stats}, which has sparked interest from the scientific community to analyse and improve their underlying security protocols. End-to-End Encryption (E2EE)~\cite{e2ee} ensures that messages can only be decrypted by their intended recipients and not by any intermediate servers. Tools for anonymity have also been expanded: Signal's sealed sender~\cite{signal-sender} allows users to send messages such that the routing servers do not learn the identity of the sender. These efforts to expand the security guarantees of instant messaging have been incorporated into widely used applications, such as WhatsApp due to its adoption of the Signal protocol~\cite{signal}. The Messaging Layer Security (MLS) protocol~\cite{mls}, recently standardised as RFC 9420, efficiently expands E2EE to group messaging.


At the same time, the service providers that own said messaging applications may desire to moderate the content shared through them, whether due to legal obligations or to prevent harassment or other forms of misconduct. Moderation becomes more difficult as the users' security guarantees increase: indeed, it is harder for service sroviders to moderate spaces in which users are anonymous or messages are deniable ---that is, it is impossible to prove that an specific user sent a given message. One solution is message franking\cite{franking}, which allows participants to prove to the service provider that some other user has acted maliciously, selectively breaking E2EE ---and for Sealed Sender, also anonymity~\cite{franking-sealed}--- to introduce moderation.

On the other hand, Anonymous Blocklisting (AB) schemes propose an alternative moderation tool that does not compromise the anonymity of users. In these schemes users interact with service providers anonymously through the use of pseudonyms derived from their identity. In order to authenticate, users provide a (zero-knowledge) proof that none of their previous pseudonyms has been blocked. While undoubtedly useful, the performance of AB schemes often scales poorly with the number of blocked users~\cite{blac, snarkblock}. Furthermore, these schemes require users to process every new insertion to the blocklist ---even if they are not affected \cite{alpaca}. As target service providers are estimated to block thousands of users per day, the efficiency of current AB schemes quickly degrades.

Both AB and message franking share in common that in order to introduce moderation, they require a powerful service provider that is able to arbitrarily enforce blocks across its user base and potentially access the contents of conversations between users. This assumption does not hold for decentralised environments in which each association of users (such as instant messaging groups) may decide their own policies for blocking users. For example, the so-called \textit{fediverse} allows every instance to define their policies, which are internally enforced by their administrators \cite{fediverse}.

In this work, we adapt the Anonymous Blocklisting framework to this environment: instead of assuming a centralised service provider with a single blocklist, our model considers a large number of \textit{Realms}, independent logical domains, each maintaining its own blocklist. Users employ different pseudonyms in each Realm, such that their activity across Realms cannot be linked. Realms can decide to enforce the blocklist of a different Realm, such that users attempting to authenticate to a Realm must also prove that they are not blocked in any of its trusted Realms. The resulting scheme is called \textit{Federated Anonymous Blocklisting} (FAB).

We provide a construction that fulfils the requirements of a FAB scheme and develop a formal security analysis to prove the security properties of Unlinkability, Blocklistability and Unframeability defined for AB schemes \cite{ab_formal}. Our construction introduces significant improvements in efficiency over existing AB schemes. Crucially, its performance scales logarithmically with the maximum size of the blocklist. While in other AB schemes users need to process every new insertion to the blocklist, even if they are not affected, our construction completely avoids that cost. However, users need to generate a proof for every realm in which the target realm trusts, introducing a different linear scaling to the performance.

The distributed nature of FAB schemes makes them ideal to an application in messaging groups, in which every group represents a Realm with its own blocklist. To prove the applicability of our proposal, we provide an implementation that incorporates FAB into the MLS protocol by allowing members to block specific pseudonyms and establish trust relationships with other groups through the use of MLS proposals. 



In summary, we present the following contributions:

\begin{itemize}
    \item A novel Federated Anonymous Blocklisting scheme. We define a system model composed of multiple Realms, each with their own blocklist. Similarly to other AB schemes, users employ unlinkable pseudonyms in different realms. In order to authenticate to a realm users must prove that they are not blocked in that realm nor in any of its trusted realms. We define the security properties of Blocklistability, Unlinkability and Non-Frameability for the FAB scheme. 
    \item A construction of the FAB scheme that employs negative accumulators and zk-SNARKs. We formally prove that the construction fulfils the FAB security properties and provide an implementation. The performance analysis shows that our implementation is more scalable than other state-of-the-art AB schemes and, crucially, does not require offline synchronisation. We demonstrate a real use case of our proposal by developing a proof-of-concept MLS variant in which every messaging group constitutes a FAB realm with its own blocklist.
\end{itemize}

The rest of this document is organised as follows. Section~\ref{sec:related} analyses related works to provide context to our contribution. In Section~\ref{sec:background} the required cryptographic primitives are introduced. Section~\ref{sec:fab} introduces and formally defines the Federated Anonymous Blocklisting scheme. Then, in Section~\ref{sec:protocol} a concrete FAB scheme is instantiated and its security is formally proven. Section~\ref{sec:impl} describes our implementation. In Section~\ref{sec:discussion} we discuss our results by comparing it to similar schemes and analysing alternative constructions. Finally, Section~\ref{sec:conclusion} will conclude this document.

\section{Related Work}
\label{sec:related}

Anonymous Blocklisting schemes allows users to authenticate to Service Providers using unlinkable pseudonyms. Service Providers also hold a blocklist of pseudonyms, such that blocked users are unable to successfully authenticate again \cite{ab_formal}. The first AB scheme was presented in  \cite{blac}; in this work, authentication costs scaled linearly with the blocklist size. Similarly, the authors of \cite{perea} present a similar scheme that limits the analysis to the most recent authentication attempts by the user. 

To improve efficiency, other AB works introduce techniques that allows users to reuse proofs. The SNARKBlock scheme \cite{snarkblock} uses aggregable zk-SNARKs to divide the blocklist into chunks such that chunk proofs could be reused. While a significant improvement, the cost of aggregating the chunk proofs still increases linearly with block size. The recent ALPACA \cite{alpaca} instead employs Incremental Verifiable Computation \cite{ivc} to achieve performance costs independent of blocklist size. Both of these works require users to perform an expensive offline synchronisation whenever new elements are inserted into the blocklist. 

Federated identity systems are composed of multiple Service Providers and Identity Providers with which users interact in order to authenticate. To provide unlinkability across various Service Providers, users employ different pseudonyms in each of them generated from the user's identity and a value provided by the Service Provider, called \textit{scoped Pseudonyms}. In  \cite{attr_pseudo}, the pseudonyms are generated from attribute credentials. The scheme defined in \cite{bison} also prevents Identity Providers from linking the user's activity by employing blind evaluation. These works allow users to be traced in a specific scope but unlinkable between scopes. However, there is no mechanism for preventing users blocked in one scope to authenticate in other scopes.

There have been recent efforts to increase anonymity and privacy in messaging groups while maintaining access control. Usually, open standards for E2EE are employed for these schemes, such as Signal \cite{signal} and Messaging Layer Security \cite{mls}. The IETF drafts \cite{mimi, mimimi} introduce \textit{metadata minimalisation} groups in which the member's credentials are encrypted and are only accessible to other group members. In \cite{aa_cgka}, a scheme for anonymously authenticating in messaging groups using Attribute-Based Credentials and fresh key pairs is defined. The authors of \cite{orca} apply Anonymous Blocklisting to protect recipients of sender-anonymous messages. Message franking \cite{franking, franking-sealed} is also employed for moderation in E2EE messaging to allow moderators to identify the original source of a message.

\section{Background}
\label{sec:background}

In this Section we introduce the cryptographic primitives that are most relevant for our protocol: zk-SNARKs and negative accumulators. Looking ahead, we will employ accumulators to model each Realm's blocklist. Users will employ zk-SNARKs to prove that none of their pseudonyms is included in any blocklist.

\subsection{zk-SNARK}

Zero Knowledge Succinct Non-interactive Arguments of Knowledge (zk-SNARK) are a type of zero-knowledge proofs characterised by their small proof size and fast verification times. These schemes are defined by a relation $R$ between a witness $w$ and a statement $x$ such that if the relation between $w$ and $x$ holds, then $(x, w) \in R$. A security parameter $\lambda$ can be derived from the description of $R$. The prover needs to reveal only $x$ to the verifier, and no information about $w$ is revealed in the proof.

A zk-SNARK scheme is composed by the following operations:

\begin{itemize}
    \item $(crs=(pk, vk), td) \gets \mathsf{Setup}(1^\lambda, R)$: Takes a relation $R$ and a security parameter $\lambda$ and outputs the Common Reference Key $crs$, which is divided into the Proving Key $pk$ and the Verification Key $vk$. Also, it outputs a trapdoor $td$.
    \item $\pi\gets \mathsf{Prove}(pk, x, w)$: From $pk$, a witness $w$ and a statement $x$, a proof $\pi$ is generated. 
    \item $Accept/Reject \gets \mathsf{Verify}(vk, x, \pi)$: From $vk$, a proof $\pi$ and the corresponding statement $x$, outputs $Accept$ or $Reject$. In the context of this work, a \textit{valid proof} refers to any $(x_v, \pi_v)$ such that $\mathsf{Verify}(vk, x_v, \pi_v) = Accept$.
    \item $\pi \gets \mathsf{Sim}(crs, td, x)$: Employs the trapdoor $td$ to simulate a proof $\pi$ for statement $x$.
\end{itemize}

A zk-SNARK scheme possesses the following security properties:

\textbf{Knowledge Soundness}. For every efficient adversary $A$, there exists an efficient extractor $Ext_A$ with access to the internal state of $A$ such that the probability 
\[
\operatorname{Pr}
\left[
\begin{array}{c}
((pk,vk), td, aux) \gets \mathsf{Gen}(R) \\
(x, \pi) \gets \mathsf{A}(R, aux, (pk, vk)) \\
w \gets Ext_A(R, aux, (pk, vk))
\\\hline
(x, w) \notin R \\
\land \mathsf{Verify}(vk, x, \pi) = Accept
\end{array}
\right]
\]
    is negligible, where $aux$ is an auxiliary input produced by $\mathsf{Gen}$. Intuitively, this means that dishonest provers could not generate a valid proof if they do not know $w$.

\textbf{Zero-Knowledge}. For every adversary A acting as a black box and $(x, w) \in R$, there exists a simulator $\mathsf{Sim}((pk, vk), aux, x)$ such that the following equality holds:
\[
\Pr\left[
\begin{array}{c}
(pk, vk), td, aux \gets \mathsf{Gen}(R) \\
\pi \gets \mathsf{Prove}(pk, x, w) \\
\hline
\mathsf{A}((pk, vk), aux, x, \pi) = 1
\end{array}
\right]
\]
$\approx$
\[
\operatorname{Pr}
\left[
\begin{array}{c}
((pk,vk), td, aux) \gets \mathsf{Gen}(R) \\
\pi \gets \mathsf{Sim}((pk, vk), td, x) \\
\hline
\mathsf{A}((pk, vk), aux, x, \pi) = 1
\end{array}
\right]
\]
where $aux$ is an auxiliary input produced by $\mathsf{Gen}$. Intuitively, this means that an attacker cannot find out anything about a witness $w$ from a proof $\pi$, a statement $x$ and a key pair $(pk, vk)$.

%
%
%


\subsection{Negative Accumulators}

A Cryptographic Accumulator is a data structure that concisely represents a set of values \cite{acc}. Alongside its benefits in space, accumulators also provide computationally efficient methods for creating and verifying (non-)membership proofs that ensure that a specific element is (or is not) in the set of values. While multiple families of accumulators exist \cite{acc_survey, acc_survey_2,acc_survey_3}, we are only interested in \textit{negative} accumulators \cite{acc_universal}. These accumulators allow for the creation on non-membership proofs, which can be used to prove that a certain element has not been inserted into it. We also require a \textit{dynamic} accumulator that allows insertions and deletions to its set of values. Formally, a negative accumulator is defined by the following operations \cite{oblivious_acc}:

\begin{itemize}
  \item $acc \gets \mathsf{Create}(1^\lambda)$: initialises an empty accumulator $acc$ with security parameter $\lambda$.
  \item $(acc', upd) \gets \mathsf{Add}(acc, x)$: adds an element $x$ to the accumulator $acc$. Returns the updated accumulator $acc'$ and an update $upd$.
    \item $(acc', upd) \gets \mathsf{Remove}(acc, x)$: removes an existing element $x$ from accumulator $acc$. Returns the updated accumulator $acc'$ and an update $upd$.
  \item $\overline{w_x}/\perp \gets \mathsf{NonMemProve}(acc, x)$: Prove non-member\-ship of $x$ in $acc$. Returns the proof $\overline{w_x}$, or $\perp$ if $x$ is in $acc$.
  \item $Accept/Reject \gets \mathsf{VerNonMem}(acc, x, \overline{w_x})$. Verifies the non-membership proof $\overline{w_x}$ for $x$ in $acc$.
\end{itemize}

Negative accumulators possess the \textit{Soundness} property, which states that for every efficient adversary $A$, the probability
\[
\operatorname{Pr}
\left[
\begin{array}{c}
acc \gets \mathsf{Create}(1^\lambda) \\
(x, \overline{w_x}) \gets A^{O_{Add}, O_{Delete}}(acc) \\
\\\hline
x \in Q \\  
\land \mathsf{VerNonMem}(acc, x, \overline{w_x}) = Accept
\end{array}
\right]
\]
is negligible, where $A$ has access to the $O_{Add}(x)$ and $O_{Delete}(x)$ queries that update $acc$ and $Q$ represents the list of current elements in the accumulator.


\section{Federated Anonymous Blocklisting}
\label{sec:fab}

In this Section, we present the framework of FAB schemes. 

\subsection{System model}

\begin{figure}
    \centering
    \includegraphics[width=0.8\linewidth]{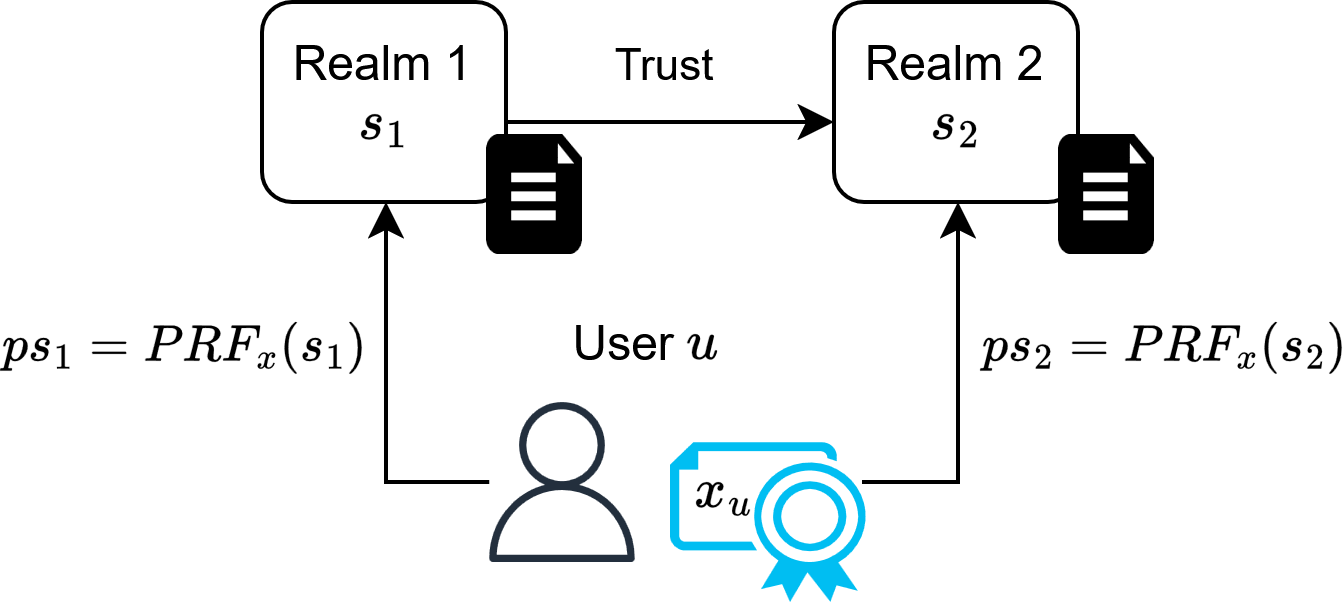}
    \caption{System model of the FAB scheme. Each Realm has a randomly-generated seed and its own blocklist, and can form trust relationships between them. Users possess a signed credential with their identity. The pseudonym of a user in a specific Realm is calculated deterministically using their respective identity and seed.}
    \label{fig:model}
\end{figure}

The system model of the FAB scheme is represented in Figure \ref{fig:model}. The following entities participate in a FAB scheme: Users ($U$) possess a private identity $x$, and their objective is to successfully authenticate to realms using a pseudonym. Realms ($R$) are abstract entities that verify the users' identity, and each realm possesses a \textit{blocklist} to which user pseudonyms can be inserted into at the realm's discretion. 

Additionally, the system requires an Identity Provider ($IP$) to create credentials for users in which they sign their identity. $IP$'s signature is trusted by the other entities in the scheme, but it is not trusted to learn the users' identity. We assume the $IP$ is able to ensure Sybil resistance, that is, users cannot possess multiple credentials referencing different identities. 

\paragraph{Authentication}

Users do not employ their identity $x$ directly to authenticate to a realm. Instead, they use a \textit{pseudonym} that is deterministically generated for every realm. In particular, realms possess a randomly-generated seed $s$ which users employ to calculate their pseudonym as $ps = \operatorname{PRF}_x(s)$. Crucially, the user's identity $x$ is private and thus pseudonyms from the same user in different realms cannot be linked.

Each realm possesses a \textit{blocklist} consisting of a list of pseudonyms. Conceptually, this blocklist represents a list of users who are considered undesirable for the realm because of previous misbehaving. Users whose pseudonym is inside a blocklist are prohibited from authenticating to that realm. Additionally, realms can dynamically establish trust relationships between each other. If a realm trusts another realm, then it also enforces its blocklist to any user attempting to authenticate to it. 

To authenticate to a realm $R$ ---henceforth referred as \textit{target realm}--- users must prove the following claims: (1) that they possess a valid credential signed by an Identity Provider, and (2) that they are not blocked in any of its \textit{trusted realms} $I$. In order to prove the latter claim, users first generate the pseudonym that corresponds to every trusted realm $\{ps^i_U \gets PRF_x(s_i)\}_{i\in I}$. Then, they prove that none of the $ps_i$ is in the blocklist in their respective realm. Since the pseudonym of an user in a specific realm can be deterministically calculated, this shows that the user has not been blocked in the trusted realm.

For practical purposes we define an auxiliary entity called \textit{Realm Directory}, which users can query to obtain public information about any realm. In practice, this Directory could forward the request to a maintainer of the realm, or consist on a server that periodically receives updates from realms.

\subsection{Comparison with Anonymous Blocklisting schemes}

In order to contextualise our work we highlight its similarities and differences with other Anonymous Blocklisting schemes \cite{ab_formal}. Both schemes operate in the same context: Users employ pseudonyms to interact with Service Providers, which can discretionarily block pseudonyms. AB's Service Providers are renamed as Realms in this work to emphasise the coexistence of an indefinite amount of them with their own context and members. Blocks in AB schemes are meant to prohibit the user from authenticating again in the same Service Provider; this also applies to our FAB scheme, but our main contribution is that it also prevents the user from authenticating in other realms.

In both approaches users derive pseudonyms from their credentials. In AB schemes a pseudonym is generated \textit{for every message} such that the actions of an user inside the same Service Provider are unlinkable. In contrast, pseudonyms in FAB are only generated during authentication and all actions by the user in a realm are performed under the same pseudonym; our scheme focuses on unlinkability \textit{across realms}. Furthermore, the pseudonyms in AB schemes are \textit{probabilistic} ---they are created with a randomly generated nonce--- whereas in FAB they are \textit{deterministic} ---each user has a specific pseudonym in each realm. The efficiency implications of this difference will be discussed later on.

\subsection{Formal Specification}

A FAB scheme consists of the following algorithms:

\begin{itemize}
  \item $crs \gets \mathsf{Setup}(1^\lambda)$: Initiates the system parameters. Outputs the Common Reference String $crs$.
  \item $cred_U \gets \mathsf{Register}(x)$. Creates a credential $cred$ signed by $IP$'s secret key for an user $U$ with identity $x$.
  \item $st \gets \mathsf{Create}(1^\lambda)$: Creates a realm with state $st$.
  \item $st' \gets \mathsf{Block}(st, ps)$: Blocks a pseudonym $ps$ in a realm. Updates the realm's state $st$.
  \item $st' \gets \mathsf{Unblock}(R, ps)$: Unblocks a pseudonym $ps$ in a realm. Updates the realm's state $st$.
  \item $(ps^R_U, \overline{w^R_U}, \pi) \gets \mathsf{Auth}(crs, cred_U, pk_{IP}, st_R,\{st_i\}_{i\in I})$: Authenticates an user $U$ with credential $cred_U$ to the target realm $R$ with the set of trusted realms $I$. Produces $U$'s pseudonym in $R$ $ps^R_U$, a proof of non-membership to $R$'s blocklist $\overline{w^R_U}$ and a proof $\pi$. Also takes as parameters $IP$'s public key $pk_{IP}$ and the realms' states $st_R, \{st_i\}_{i\in I}$.
  \item $Accept/Reject \gets \mathsf{Verify}(crs, pk_{IP}, ps^R, \overline{w^R}, \pi, st_R,$ $\{st_i\}_{i\in I})$. Verifies the proof $\pi$ for target realm $R$ with the set of trusted realms $I$. Also takes as parameters $IP$'s public key $pk_{IP}$, a pseudonym and proof of non-membership $ps^R, \overline{w^R}$ and the realms' states $st_R, \{st_i\}_{i\in I}$.
\end{itemize}

We remark that the $\mathsf{Auth}$ and $\mathsf{Verify}$ algorithms take as parameter the set of trusted realms for the target realm. Thus, the establishment of trust relationships between realms is handled outside of the FAB scheme and it is up to the callers of both algorithms to use the correct realms. 

\begin{figure}
\centering
\includegraphics[width=\linewidth]{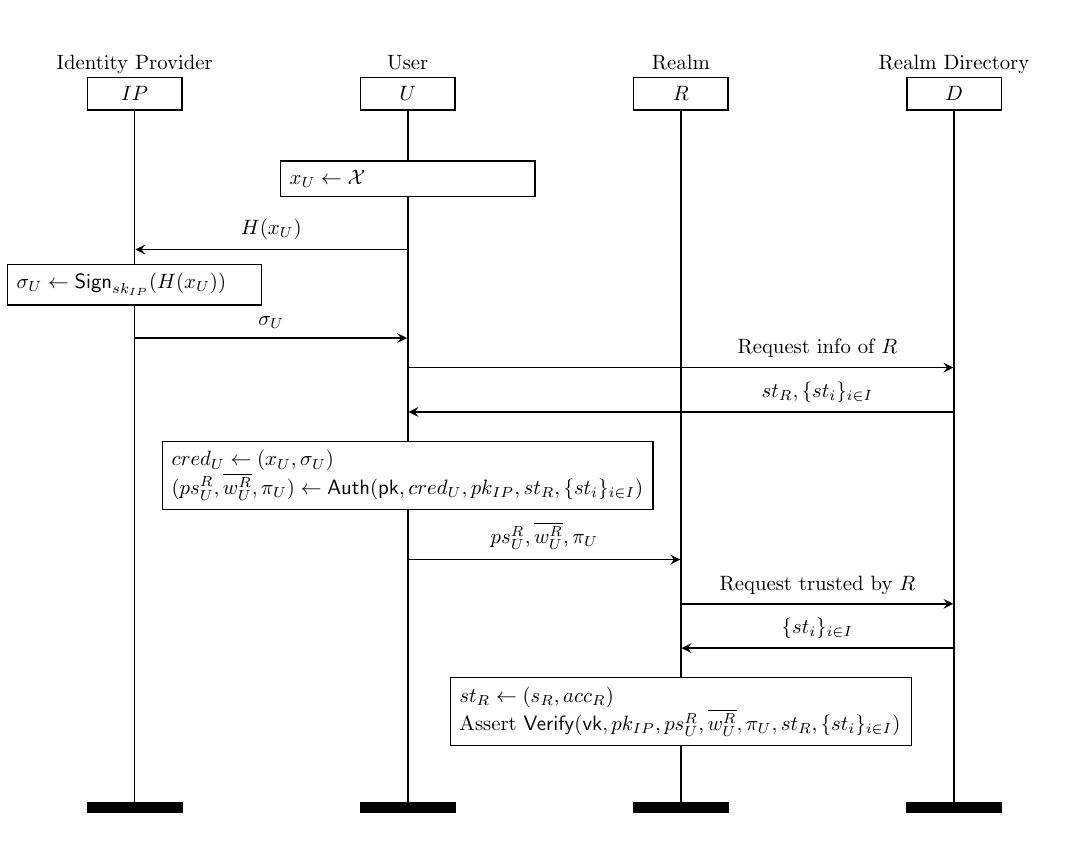}
\caption{Authentication flow of the FAB scheme. The User $U$ first registers its identity with the Identity Provider and then obtains the list of realms trusted by $R$. $U$ then executes $\mathsf{Auth}$ to prove they are not blocked, which is later verified by $R$.}
\label{fig:flow}
\end{figure}

Figure~\ref{fig:flow} shows an example execution flow of a FAB scheme. First, an user $U$ registers their identity to $IP$ and receives a signed credential. In order to authenticate to target realm $R$, $U$ must first obtain from the Realm Directory $R$'s state $st_R$ and the state of all of its trusted realms $\{st_i\}_{i\in I}$. Then, $U$ executes the $\mathsf{Auth}$ algorithm to create a pseudonym $ps^R_U$, a proof of non-membership to $R$'s blocklist $\overline{w^R_U}$ and a proof $\pi$. When $R$ receives the authentication attempt, it first retrieves the state of its trusted realms to obtain the latest updates. Finally, it executes the $\mathsf{Verify}$ algorithm validate $\pi$.

\subsection{Security}

Our security analysis follows the standard Anonymous Blocklisting framework of \cite{blac, snarkblock}:  we also employ the security notions of Blocklistability, Unlinkability and Non-Frameability. However, we provide custom definitions of said properties to account for the differences in our system model.

We follow the query-based approach of \cite{alpaca}. The security properties are defined as a game between a Challenger and an Adversary. Each security game has its own winning condition and may incorporate checks to prevent trivial wins by the adversary. Figure~\ref{alg:shared} shows the shared queries between all security games ---with a slight variation in the $Q_{Block}$ query for the Non-Frameability and Unlinkability games. They are mostly related to administrative tasks such as creating realms, registering users and blocking and unblocking pseudonyms. 

\begin{figure}
\small
\begin{multicols}{2}
\begin{algorithmic}

\Function{$Q_{Create}$}()
\State $st_R \gets \mathsf{Create}(1^\lambda)$
\State $R[realms] \gets st_R$
\State $realms \gets realms + 1$
\State \Return $(realms, st_R)$
\EndFunction

\\

\Function{$Q_{Register}$}{$x$}
\State $cred_x \gets \mathsf{Register}(x)$
\State $C[creds] \gets C \cup \{cred_x\}$
\State $creds \gets creds + 1$
\State \Return $(creds, cred_x)$
\EndFunction

\\

\Function{$Q_{Block}$}{$\textcolor{blue}{x, }r, ps$}
\State $st_R \gets R[r]$
\textcolor{red}{\Assert{ $r \neq r_c$}}
\textcolor{blue}{\Assert{ $is\_pseudonym(x, st_R, ps)$}}
\State $st'_R \gets \mathsf{Block}(st_R, ps)$
\State $B \gets B \cup \{(\textcolor{blue}{x, }r, ps)\}$
\State $R[r] \gets st'_R$
\State \Return $(st'_R)$
\EndFunction

\columnbreak
\Function{$Q_{Unblock}$}{$r, ps$}
\State $st_R \gets R[r]$
\State $st'_R \gets \mathsf{Unblock}(st_R, ps)$
\State $B \gets B \setminus \{(r, ps)\}$
\State $R[r] \gets st'_R$
\State \Return $(st'_R)$
\EndFunction

\\

\Function{$is\_pseudonym$}{$x, st, ps$}
\State $(s, acc) \gets st$
\State \Return $PRF_x(s) = ps$
\EndFunction
\end{algorithmic}
\end{multicols}
\caption{Shared queries and auxiliary functions through all security games. Blue-highlighted text in the \textit{Block} query only applies to the Non-Frameability game. Red-highlighted text applies only to the Unlinkability game.}
\label{alg:shared}
\end{figure}


\begin{figure}
\small
\begin{algorithmic}

\Function{BL}{$\lambda$}
\State $(\mathsf{pk}, \mathsf{vk}) \gets \mathsf{Setup(1^\lambda}, R^{auth}, R^{ban})$
\State $C, R \gets \varnothing; creds, realms \gets 0$
\State $(r, I, ps, \overline{w^R}, \pi) \gets \mathcal{A}^Q(\mathsf{pk})$
\State $st_R \gets R[r]$
\State $\{st_i\}_{i\in I} \gets R[I]$
\State $all\_blocked \gets \bigwedge_{cred\in C} (blocked(r, cred) \lor \bigvee_{i\in I} blocked(i, cred))$
\State \Return $\mathsf{Verify}(\mathsf{vk}, st_R,  pk_{ip}, ps, \overline{w^R}, \pi,\{st_i\}_{i\in I}) = Accept \land all\_blocked$
\EndFunction

\\

\Function{$blocked$}{$i, cred$}
\State $st_i \gets R[i]$
\State \Return $\exists ps$ $s.t.$ $(i, ps)\in B \land is\_pseudonym(cred.x, st_i, ps)$ 
\EndFunction

\end{algorithmic}
\caption{Blocklistability Security game.}
\label{alg:bl}
\end{figure}


\begin{figure}
\small
\begin{algorithmic}

\Function{Unlink}{$\lambda$}
\State $(\mathsf{pk}, \mathsf{vk}) \gets \mathsf{Setup(1^\lambda}, R^{auth}, R^{ban})$
\State $(pk_{ip}, sk_{id}) \gets \mathsf{IP.Init}(1^\lambda)$
\State $A, R, B \gets \varnothing; creds, realms \gets 0$
\State $b \overset{{\scriptscriptstyle \operatorname{R}}}{\gets} \{0,1\}$
\For{$i$ in \{0,1\}}
    \State $cred_i\gets \mathsf{Register}(1^\lambda)$
\EndFor
\State $b' \gets \mathcal{A}^Q(\mathsf{pk}, pk_{ip})$
\State \Return $b=b'$
\EndFunction

\\
   
\Function{$Q_{Chal}$}{$r_c, I$}
\State $st_R \gets R[r_c]$
\State $\{st_i\}_{i\in I} \gets R[I]$
\Assert{ $r_c \notin A$}
\For{$u$ in \{0,1\}}
    \Assert{ $\nexists ps$ $s.t.$ $(r_c, ps)\in B \land is\_pseudonym(cred_u.x, st_R, ps)$}
    \Assert{ $\nexists ps$ $s.t.$ $(i, ps)\in B \land is\_pseudonym(cred_u.x, st_i, ps)$ for $i\in I$}
\EndFor
\State $(ps^R_b, \overline{w^R_b}, \pi_b) \gets \mathsf{Auth}(\mathsf{pk}, cred_b, pk_{ip}, st_R, \{st_i\}_{i\in I})$
\State \Return $(ps^R_b, \overline{w^R_b}, \pi_b)$
\EndFunction

\\

\Function{$Q_{Auth}$}{$u, r, I$}
\Assert{ $r \neq r_c$}
\State $st_R \gets R[r]$
\State $\{st_i\}_{i\in I} \gets R[I]$
\State $(ps^R_u, \overline{w^R_u}, \pi_u) \gets \mathsf{Auth}(\mathsf{pk}, cred_u, pk_{ip}, st_R, \{st_i\}_{i\in I})$
\State $A \gets A\cup \{r\}$
\State \Return $(ps^R_u, \overline{w^R_u}, \pi_u)$
\EndFunction

\end{algorithmic}
\caption{Unlinkability Security game.}
\label{alg:unlink}
\end{figure}


\begin{figure}
\small
\begin{algorithmic}

\Function{NF}{$\lambda$}
\State $(\mathsf{pk}, \mathsf{vk}) \gets \mathsf{Setup(1^\lambda}, R^{auth}, R^{ban})$
\State $(pk_{ip}, sk_{id}) \gets \mathsf{IP.Init}(1^\lambda)$
\State $C, R, B \gets \varnothing; creds, realms \gets 0$
\State $(u, r, I) \gets \mathcal{A}^Q(\mathsf{pk}, pk_{ip})$
\State $cred_u \gets C[u]$
\State $st_R \gets R[r]$
\State $\{st_i\}_{i\in I} \gets R[I]$
\State $(ps^R_u, \overline{w^R_u}, \pi_u) \gets \mathsf{Auth}(\mathsf{pk}, cred_u, pk_{ip}, st_R, \{st_i\}_{i\in I})$
\State \Return $\mathsf{Verify}(\mathsf{vk}, st_R,  pk_{ip}, ps^R_u, \overline{w^R_u}, \pi_u,\{st_i\}_{i\in I}) = Reject \land (cred_u.x, r) \notin B \land \forall i \in I$ $(cred_U.x, i) \notin B$
\EndFunction

\end{algorithmic}
\caption{Non-Frameability Security game.}
\label{alg:nf}
\end{figure}

\subsubsection{Blocklistability}

The Blocklistability property states that a successful authentication can only be generated by a registered user that is not blocklisted neither in the target realm nor in any of its trusted realms. Thus, this property also considers authentication attempts by unregistered users.

The security game is shown in Figure~\ref{alg:bl}, and it unfolds as follows: the Challenger initialises the environment and allows the adversary to execute arbitrary queries. Eventually, the adversary outputs a target realm, a set of trusted realms, and an authentication attempt. The adversary wins if said authentication attempt is valid but all created users are blocked on either the target realm or on at least one of the trusted realms.

We define $A$'s advantage in breaking Blocklistability as 
\begin{equation}
  \label{eq:adv-bl}
    \mathsf{Adv}^{BL}_A(\lambda) = \Pr [\text{$A$ wins} ]
\end{equation}

\begin{definition}
A FAB scheme provides Blocklistability if for all efficient adversaries $A$, $\mathsf{Adv}^{BL}_A(\lambda)$ is negligible.
\end{definition}

\subsubsection{Unlinkability}

The Unlinkability property ensures that it is impossible to identify if two pseudonyms in different realms correspond to the same user without knowing the user's original credential.

The security game is shown in Figure\ref{alg:unlink}, and it unfolds as follows: the Challenger creates two users with credentials $cred_0, cred_1$, and the adversary is allowed to execute arbitrary queries. Eventually the adversary queries $Q_{Chal}$ with references to a realm and a set of trusted realms. The Challenger then authenticates $cred_b$ in said realm. The adversary wins if they successfully guess $b$. The $Q_{Auth}$ query allows the adversary to obtain honestly-generated authentication attempts from the challenge credentials, as the adversary cannot generate them by themselves. Trivial attacks are prevented by the following checks: (1) querying $Q_{Chal}$ for a realm in which either user would not be able to generate a valid proof or (2) the adversary already knows the pseudonym of a user, (3) querying $Q_{Auth}$ for the challenge realm or (4) blocking the challenge pseudonym, which would provoke future $Q_{Auth}$ queries to fail only for the challenge user. The latter check is highlighted in green in Figure~\ref{alg:shared}.

We define $A$'s advantage in breaking Unlinkability as 
\begin{equation}
  \label{eq:adv-unlink}
    \mathsf{Adv}^{Unlink}_A(\lambda) = \bigl| \Pr [b=b'] -\frac{1}{2} \bigr|.
\end{equation}

\begin{definition}
A FAB scheme provides Unlinkability if for all efficient adversaries $A$, $\mathsf{Adv}^{Unlink}_A(\lambda)$ is negligible.
\end{definition}

\subsubsection{Non-frameability}

The Non-frameability property states that it is impossible to prevent a non-blocklisted user from authenticating.

The security game is shown in Figure \ref{alg:nf}, and it unfolds as follows: the Challenger initialises the environment and allows the adversary to execute arbitrary queries. Eventually, the adversary selects one of the registered users, a target realm and a set of trusted realms. The Challenger then attempts to authenticate the user. The adversary wins if the user has not been blocked but the authentication fails. 

As shown in Figure~\ref{alg:shared}, the $O_{Block}$ query is different for the Non-frameability security game, as it also requires the adversary to submit an user's secret value $x$ \cite{alpaca}. This allows the Challenger to identify which user is being targeted in order to later check if the target user has been blocklisted.

We define $A$'s advantage in breaking Non-frameability as 
\begin{equation}
  \label{eq:adv-nf}
    \mathsf{Adv}^{NF}_A(\lambda) = \Pr [\text{$A$ wins} ].
\end{equation}

\begin{definition}
A FAB scheme provides Non-frameability if for all efficient adversaries $A$, $\mathsf{Adv}^{NF}_A(\lambda)$ is negligible.
\end{definition}

\section{Proposed FAB Scheme}
\label{sec:protocol}

We now provide a concrete construction of a FAB scheme. The proposed FAB scheme uses as primitives a pseudo-random function $PRF$, a hash function $H$, a signature scheme $S = \{\mathsf{S.KeyGen}, \mathsf{S.Sign}, \mathsf{S.Verify}\}$, a negative accumulator $AC = \{\mathsf{AC.Create}, \mathsf{AC.Add}, \mathsf{AC.Remove}, \mathsf{AC.NonMemProve}, \allowbreak \mathsf{AC.VerNonMem}\}$ and a zk-SNARK scheme $ZK = \{\mathsf{ZK.Gen}, \mathsf{ZK.Sign}, \mathsf{ZK.Verify}\}$. We refer to a user as $U$, a target realm as $R$ and a trusted realm as $R'$. 

\begin{figure}
\small
\begin{algorithmic}

\Function{Setup}{$1^\lambda, R^{auth}, R^{block}$}
\State $(pk^{auth}, vk^{auth}) \gets \mathsf{ZK.Setup}(1^\lambda, R^{auth})$
\State $(pk^{block}, vk^{block}) \gets \mathsf{ZK.Setup}(1^\lambda, R^{block})$
\State $\mathsf{pk} \gets (pk^{auth}, pk^{block})$
\State $\mathsf{vk} \gets (vk^{auth}, vk^{block})$
\State \Return $(\mathsf{pk}, \mathsf{vk})$ 
\EndFunction

\begin{multicols}{2}

\Function{Register}{$x$}
\State $\sigma \gets IP.Sign(H(x))$
\State $cred \gets (x, \sigma)$
\State \Return $cred$
\EndFunction

\columnbreak

\Function{Create}{$1^\lambda$}
\State $s_R \gets \mathcal{X}$
\State $acc_R \gets \mathsf{AC.Create}(1^\lambda)$
\State $st_R \gets (s_R, acc_R)$
\State \Return $st_R$ 
\EndFunction

\end{multicols}

\end{algorithmic}
\caption{Initialisation Functions of the FAB scheme.}
\label{alg:init}
\end{figure}

\begin{figure}
\begin{multicols}{2}
\small
\begin{algorithmic}

\Function{Block}{$st_R, ps$}
\State $(s_R, acc_R) \gets st_R$
\State $acc'_R \gets \mathsf{AC.Add}(acc'_R, ps)$
\State \Return $(s_R, acc'_R)$
\EndFunction

\columnbreak

\Function{Unblock}{$st_R, ps$}
\State $(s_R, acc_R) \gets st_R$
\State $acc'_R \gets \mathsf{AC.Delete}(acc'_R, ps)$
\State \Return $(s_R, acc'_R)$
\EndFunction

\end{algorithmic}
\end{multicols}
\caption{Realm functions.}
\label{alg:realm}
\end{figure}

\begin{figure}
\small
\begin{algorithmic}

\Function{Auth}{$\mathsf{pk}, cred_U, pk_{IP}, st_R, \{st_i\}_{i\in I}$}
\State $(x_U, \sigma_U) \gets cred_U$
\State $(pk^{auth}, pk^{block}) \gets \mathsf{pk}$
\State $(s_R, acc_R) \gets st_R$
\State $ps^R_U \gets PRF_{x_U}(s_R)$
\State $\overline{w^R_U} \gets \mathsf{AC.NonMemProve}(acc_R, ps^R_U)$
\State $\pi^{auth} \gets \mathsf{ZK.Prove}(pk^{auth}, (pk_{IP}, ps^R_U, s_R), (\sigma_U, x_U))$
\For{$i\in I$}
    \State $(s_i, acc_i) \gets st_i$
    \State $ps^i_U \gets PRF_{x_U}(s_i)$
    \State $\overline{w^i_U} \gets \mathsf{AC.NonMemProve}(acc_i, ps^i_U)$
    \State $\pi^{block}_i \gets \mathsf{ZK.Prove}(pk^{block}, (ps, s_R, s_i, acc_i), (\overline{w^i_U}, x_U))$
\EndFor
\State $\pi_U \gets (\pi^{auth}, \{\pi^{block}_i\}_{i\in I})$
\State \Return $(ps^R_U, \overline{w^R_U}, \pi_U)$
\EndFunction

\\

\Function{Verify}{$\mathsf{vk}, pk_{IP}, ps^R, \overline{w^R}, \pi, st_R, \{st_i\}_{i\in I}$}
\State $(vk^{auth}, vk^{block}) \gets \mathsf{vk}$
\State $(s_R, acc_R)\gets st_R$
\State $(\pi^{auth}, \{\pi^{block}_i\}_{i\in I}) \gets \pi$
\Assert { $\mathsf{AC.VerNonMem}(acc_R, ps^R, \overline{w^R})$}
\Assert{ $\mathsf{ZK.Verify}(vk^{auth}, (pk_{IP}, ps, s_R), \pi^{auth})$}
\For{$i \in I$}
    \State $(s_i, acc_i)\gets st_i$
    \Assert{ $\mathsf{ZK.Verify}(vk^{block}, (ps, s_R, s_i, acc_i), \pi^{block}_i)$}
\EndFor
\EndFunction

\end{algorithmic}
\caption{Authentication and Verification Functions.}
\label{alg:auth}
\end{figure}

\begin{figure}

\begin{multicols}{2}
\small
\begin{algorithmic}

\Function{Init}{$1^\lambda$}
\State $(pk_{IP}, sk_{IP}) \gets \mathsf{S.KeyGen}(1^\lambda)$
\State \Return $(pk_{IP}, sk_{IP})$
\EndFunction

\columnbreak

\Function{Sign}{$sk_{IP}, h_x$}
\State $\sigma_x \gets \mathsf{S.Sign}(h_x)$
\State \Return{$\sigma_x$}
\EndFunction

\end{algorithmic}
\end{multicols}
\caption{Identity Provider Functions.}
\label{alg:ip}
\end{figure}

\textbf{State.} New realms and users are respectively initialised through the algorithms $\mathsf{Register}$ and $\mathsf{Create}$, shown in Figure~\ref{alg:init}. The realms's state $st_R$ is composed of two elements: a seed $s_R$ employed by users to create their pseudonym, and a negative accumulator $acc_R$ representing the realm's blocklist. The users' state consists on a credential $cred_U$ that contains the user's identity $x$ and $IP$'s signature over $H(x)$.

The instantiation of algorithms $\mathsf{Block}$ and $\mathsf{Unblock}$ are shown in Figure~\ref{alg:realm}. They are simply tasked with adding or removing elements from the realm's accumulator.

\textbf{zk-SNARK Relations.} In order to successfully authenticate to a realm, an user must prove (1) that they possess a valid credential signed by the Identity Provider and (2) that they are not blocked in any of the trusted realms. We respectively formalise the two statements as the relations $R^{auth}$ and $R^{block}$, to be proven through a zk-SNARK. Figure~\ref{fig:relation} shows the description of both relations. 

The relation $R^{auth}$ is shown in Equation~\ref{eq:relation-auth}: its statement is the issuer's public key $pk_{IP}$, the target realm's seed $s_R$ and the user's pseudonym in the target realm $ps^R_U$, and its witness is the user's credential $(x_U, \sigma_U)$. $R^{auth}$ verifies that the user's pseudonym is correctly computed from their identity and the realm's seed, and that the user's credential is adequately signed by the Identity Provider. We refer to the zk-SNARK proof that proves the $R^{auth}$ relation as $\pi^{auth}$.

\begin{figure}
\centering

\begin{tcolorbox}[colback=white,colframe=black, width=\columnwidth]
\begin{equation}
  \label{eq:relation-auth}
  R^{auth} = \left\{
  \begin{array}{l}
    \begin{array}{@{}l@{}}
    ((pk_{IP}, s_R, ps^R_U), (x_U, \sigma_U))
    \end{array}
    \\
    \hline
    \begin{array}{@{}l@{}}
      \mathsf{S.Verify_{pk_{IP}}(\sigma_U, x_U)}, \\
      ps^R_U = PRF_{x_U}(s_R)
    \end{array}
    \end{array}
  \right\}
\end{equation}

\begin{equation}
  \label{eq:relation-block}
  R^{block} = \left\{
  \begin{array}{l}
    \begin{array}{@{}l@{}}
    ((ps^R_U, s_R, s_{R'}, acc_{R'}), (\overline{w^{R'}_x}, x_U))
    \end{array}
    \\
    \hline
    \begin{array}{@{}l@{}}
      ps^R_U = PRF_{x_U}(s_R) \\
      ps^{R'}_U = PRF_{x_U}(s_{R'}) \\
      \mathsf{AC.VerNonMem}(acc_{R'}, ps^{R'}_U, \overline{w^{R'}_x}), \\
    \end{array}
    \end{array}
  \right\}
\end{equation}
\end{tcolorbox}

\caption{Description of the relations $R^{auth}$ and $R^{block}$.}
\label{fig:relation}
\end{figure}

Equation~\ref{eq:relation-block} shows the $R^{block}$ relation. Its statement is the user's pseudonym in the target realm $ps^R_U$, the seeds of both the target and trusted realm $(s_R, s_{R'})$, and the accumulator of the target realm $acc_{R'}$. Its witness is the user's identity $x_U$ and a proof of non-membership to the target realm's accumulator $\overline{w^{R'}_x}$. The relation mainly checks that the user is not blocked in the trusted realm: to that end, the pseudonym the user \textit{would have in the trusted realm} $ps^{R'}_U$ is computed. Then, the proof of non-membership for that pseudonym is verified. We note that the $R^{block}$ relation also verifies that the user's pseudonym fo the target realm is computed correctly. This may seem redundant, as $R^{auth}$ also verified the same condition. However, it is necessary to ensure that the same identity $x_U$ was used for both relations. We refer to the zk-SNARK proof that proves the $R^{block}$ relation as $\pi^{block}$.

The $\mathsf{Setup}$ algorithm, shown in Figure~\ref{alg:init}. $\mathsf{Setup}$ is executed by a trusted party such as the Identity Provider, and it initialises the zk-SNARK public parameters for both relations.

The $\mathsf{Auth}$ instantiation is shown in Figure~\ref{alg:auth}. It involves the creation of $U$'s pseudonym for target realm $ps^R_U$ as well as for every trusted realm $ps^i_U$. A zk-SNARK proof for the $R^{auth}$ is created using the target realm's seed. Then, a zk-SNARK proof of the $R^{block}$ relation is generated for all trusted realms' blocklists, using the corresponding user pseudonym. Thus, a total of $i+1$ zk-SNARK proofs are generated, where $i$ is the number of trusted realms.

The $\mathsf{Verify}$ instantiation, also shown in Figure\ref{alg:auth}, follows a similar execution flow. All zk-SNARK proofs generated in the $\mathsf{Auth}$ algorithm are verified, as well as the non-membership proof for the pseudonym $ps^R$. 

\subsection{Security}

We now prove that the proposed scheme provides the security properties defined for FAB schemes in Section \ref{sec:fab}.

\begin{theorem}
\label{th:bl}
If $ZK$ provides Knowledge Soundness, $AC$ provides Soundness, $PRF$ is a secure pseudo-random function and $S$ is EUF-CMA secure, then the proposed FAB scheme provides Blocklistability.
\end{theorem}

\begin{proof}
We start by applying a series of transforms to the security game:

\begin{itemize}
    \item $G_1$ is identical to the Blocklistability game described in Figure \ref{alg:unlink}, but the Challenger executes an extractor $E$ for the \textit{Auth} relation, obtaining $(\sigma, x^{auth}) \gets \mathcal{E}(\pi^{auth})$. Since the zk-SNARK scheme is knowledge-sound, the probability that $ps \neq \operatorname{PRF}_{x^{auth}}(s_R)$ or $\mathsf{S.Verify_{pk_{IP}}}(\sigma, H(x)) = Reject$ is negligible.
    \item $G_2$ is identical to $G_1$ but the Challenger executes an extractor $E$ for the $R^{block}$ relation, obtaining $(\overline{w^i}, x^{block}_i) \gets \mathcal{E}(\pi^{block}_i)$ for every $i\in I$. Since the zk-SNARK scheme is knowledge-sound, the probability that $ps \neq \operatorname{PRF}_{x^{block}_i}(s_R)$ or $\mathsf{AC.VerNonMem}(acc_i, \operatorname{PRF}_{x^{block}_i}(s_i), \overline{w^i}) = Reject$ for any $i\in I$ is negligible.
\end{itemize}

We have established that $\operatorname{PRF}_{x^{auth}}(s_R) = PRF_{x^{block}_i}(s_R) = ps$. Since $PRF$ is collision-resistant, the probability of $x^{auth} \neq x^{block}_i$ is negligible. Thus, we will refer to them simply as $x$.

Let $B$ be an adversary playing an EUF-CMA game for the signature scheme $S$, that acts as a wrapper for $A$. 
During $Q_{Register}$ queries, $B$ requests its challenger for a signature of $H(x)$. Eventually, $A$ outputs $(r, I, ps, \overline{w^R}, \pi)$ and $B$ obtains $(\sigma, x^{auth})$ from $E$. $B$ then submits $(\sigma, H(x))$ to its challenger. Clearly, if $A$ was able to forge $\sigma$ then $B$ would win its game. Thus, the chance that $A$ successfully uses a $\sigma$ not obtained through the $O_{Register}$ query is negligible. Let $cred$ be the credential generated through a $Q_{Register}$ query that contains $(\sigma, x)$. 

Let $B$ be an adversary playing simultaneous Soundness games for the accumulator $AC$, that acts as a wrapper for $A$. 
Whenever $A$ calls $Q_{Create}$, $B$ initiates a game with a new challenger and forwards the received $acc$. During $O_{Block}$ and $O_{Unblock}$ queries, $B$ will request the corresponding challenger to add and delete the received $ps$. Eventually, $A$ outputs $(r, I, ps, \overline{w^R}, \pi)$ and $B$ obtains $\{(\overline{w^i}, x)\}_{i\in I}$ from $E$. $B$ calculates $\{ps_i \gets PRF_x(s_i)\}_{i\in I}$ and submits $(ps, \overline{w^R}) \cup \{(ps_i, \overline{w^i})\}$ to each corresponding challenger. 

Let $j \in I\cup\{r\}$ such that $(j, ps_j) \in B$ --- as specified by the $all\_blocked$ predicate, at least one such $j$ exists. That means the $j$-th challenger has been queried with $ps_j$. If $A$ is able to produce a $(ps_j, \overline{w^j})$ such that $\mathsf{AC.Verify}(acc_j, ps_j, \overline{w^j}) = Accept$, then $B$ would win its game against the $j$-th challenger. Thus, the probability of $A$ successfully forging a non-membership proof for a blocked pseudonym is negligible.
\end{proof}

\begin{theorem}
\label{th:unlink}
If $PRF$ is a secure pseudo-random function and $ZK$ provides Zero-Knowledge, then the proposed FAB scheme provides Unlinkability.
\end{theorem}

\begin{proof}
We start by applying the following transform to the security game: $G_1$ is identical to the Unlinkability game described in Figure \ref{alg:unlink}, but every call to $\mathsf{ZK.Prove}(pk, x, w)$ is replaced by $\mathsf{ZK.Sim}(crs, td, x)$. As stated by the Zero-Knowledge property of the zk-SNARK scheme, $G_2$ is indistinguishable from the original game. 

Recall that the output of $Q_{Chal}$ is $(ps^R_b, \overline{w^R_b}, (\pi^{auth}_b, \{\pi^{block}_i\}_{i\in I}))$. As per $G_1$, none of these values reveals anything about $cred_b$: $ps^R_b$ is indistinguishable from random and $\pi^{auth}_b$ and $\{\pi^{block}_i\}_{i\in I})$ are simulated proofs. The latter point implies that none of the values of $cred_b = (x_b, \sigma_b)$ were used in the generation of the proofs. The same logic applies to any $(ps^R_u, \overline{w^R_u}, \pi_u)$ output by $Q_{Auth}$, as its zk-SNARK proofs are also simulated. 

We now show that $A$ cannot link two pseudonyms in different realms to the same $x_b$. Let $B$ be an adversary playing a Key Indistinguishability game for the pseudo-random function $\operatorname{PRF}$, that acts as a wrapper for $A$. 
Whenever $B$ would execute $PRF$, it instead queries its challenger. $B$ does not generate $cred_0,cred_1$ as they are not needed for the simulated zk-SNARK proofs. Eventually, $A$ outputs $b'$ and $B$ forwards it to its challenger. Clearly, if $A$ succeeds, then $B$ would also win the Key Indistinguishability game. As $\operatorname{PRF}$ is assumed to be secure, $A$ cannot obtain any information about the $x_b$ used and thus wins with negligible probability.
\end{proof}

\begin{theorem}
\label{th:nf}
If $\operatorname{PRF}$ is a secure pseudo-random function and $AC$ is a correct accumulator, then the proposed FAB scheme provides Non-frameability.
\end{theorem}

\begin{proof}
The values $(ps^R_u, \overline{w^R_u}, \pi_u$) are honestly generated by the Challenger. Thus, the Correctness property of the accumulator ensures that if $ps^R_u$ has not been added to $acc_R$, then $\mathsf{AC.VerNonMem}(acc_R, ps^R_u, \overline{w^R_u})$ will accept. Likewise, the Completeness property of the zk-SNARK scheme ensures that if neither $ps^i_u = \operatorname{PRF}_{cred_u.x}(s_i)$ for $i\in I$ has been added to $acc_i$, then $\mathsf{ZK.Verify}(vk^{block}, \pi^{block}_i, (ps^R_u, s_R, s_i, acc_i))$ will accept. The same applies to the verification of $R^{auth}$.

Thus, $A$ can only win by finding and blocking a $ps^i_y = \operatorname{PRF}_y(s_i)$ s.t. $ps^i_y = ps^i_u = PRF_{cred_u.x}(s_i)$, for any $i\in I \cup {r}$. Clearly, one of the checks in $\mathsf{Verify}$ would reject: either $\mathsf{AC.VerNonMem}(acc_R, ps^R_u, \overline{w^R_u})$ - if $i = r$ - or $\mathsf{ZK.Verify}(vk^{block}, \pi^{block}_i, (ps^R_u, s_R, s_i, acc_i))$ - if $i \in I$. However, $PRF$ is collision-resistant and thus the chance of finding a collision is negligible. 
\end{proof}

\section{Implementation}
\label{sec:impl}

We have implemented the proposed FAB scheme in $\sim3000$ lines of Rust code.\footnote{Available at \url{https://github.com/SDABIS/mls_fab}} Table~\ref{tab:ops} shows the concrete algorithms employed for each of the cryptographic primitives our construction depends on. All zk-SNARK operations are handled by the \textit{arkworks} ecosystem~\cite{arkworks}.

\begin{table}[tb]
\caption{Algorithms employed for the implementation.}
\centering
\begin{tabular}{@{}ll@{}}
\toprule
Operation       & Algorithm                \\
\midrule
Signature       & Schnorr (JubJub Curve)  \\
Hash Function   & Poseidon \cite{poseidon}\\
PRF             & Poseidon\\
zk-SNARK curve  & BLS12-381                \\
zk-SNARK scheme & Groth16 \cite{groth16} \\
\bottomrule
\end{tabular}
\label{tab:ops}
\end{table}

%


In our implementation we employ a Merkle Tree as negative accumulator to represent the list of banned users. This structure allows for logarithmic complexity in membership verification, which is particularly relevant as verification is executed inside the \texttt{Block} circuit~\cite{zcash}. Merkle Trees only allow for membership proofs, whereas we only require non-membership proofs. To that end, we construct a \textit{Complementary Merkle Tree}~\cite{promises}: starting with the set of all possible pseudonyms, it is divided into intervals $[a,b)$ with a breakpoint at all of the banned pseudonyms. Every leaf in the Complementary Merkle Tree represents one of said intervals and contains $H(a||b)$. To prove non-membership of $ps$, users must prove that there exists a leaf at position $i$ with content $H(a||b)$ such that $a \leq ps < b$, that is, $\overline{w_x} = (a, b, i, ps)$.

We additionally include in the implementation a Realm Directory as described in Figure \ref{fig:flow}, instantiated as a REST API web server. Realms upload their states to the Directory, which are downloaded by clients for the authentication. 

\subsection{Integration into MLS}

We also provide a concrete application of the theoretical concepts of FAB to demonstrate its real-world usefulness. In particular, we apply FAB to the context of instant messaging groups. Each realm represents a different messaging group with its own blocklist. We employ the Messaging Layer Security~\cite{mls} framework for our implementation.

Every FAB operation is expressed in the terms of MLS. The functions that modify the state of the realm ---$\mathsf{Block}$, $\mathsf{Unblock}$ and establishing trust relations--- are performed by the \textit{proposal-and-commit} paradigm: users can propose this modifications to the group, which are later committed.

We make use of the extensibility options provided by MLS to implement our FAB scheme. The realm state is inserted into the Group Context, such that any group member has access to it and can be published to external users attempting to join the group. New members present the authentication proof generated by the $\mathsf{Auth}$ function in their \textit{KeyPackage}, a structure defined by MLS that also contains cryptographic information required to join a group. Current members verify 

We note that establishing consensus among group members about which pseudonyms should be added is outside the scope of this work. There exist other works that are concerned with authorisation in MLS groups that are compatible with our environment~\cite{cgka_fa, a_cgka}.

\subsection{Evaluation}
\label{sec:eval}

In this Section, we evaluate the performance of our proposed FAB scheme. For comparison, benchmarks for the state-of-the-art SNARKBlock \cite{snarkblock} and ALPACA \cite{alpaca} schemes are also included. SNARKBlock is configured with chunk sizes of 1024 pseudonyms and a buffer of 14 chunks of size 16. The measurements for all three schemes were obtained in the same execution environment, using the code provided in their respective repositories~\cite{snarkblock-impl, alpaca-impl}.

\begin{figure*}[t]
    \centering
    \begin{subfigure}[b]{0.32\textwidth}
        \centering
        \includegraphics[width=1\linewidth]{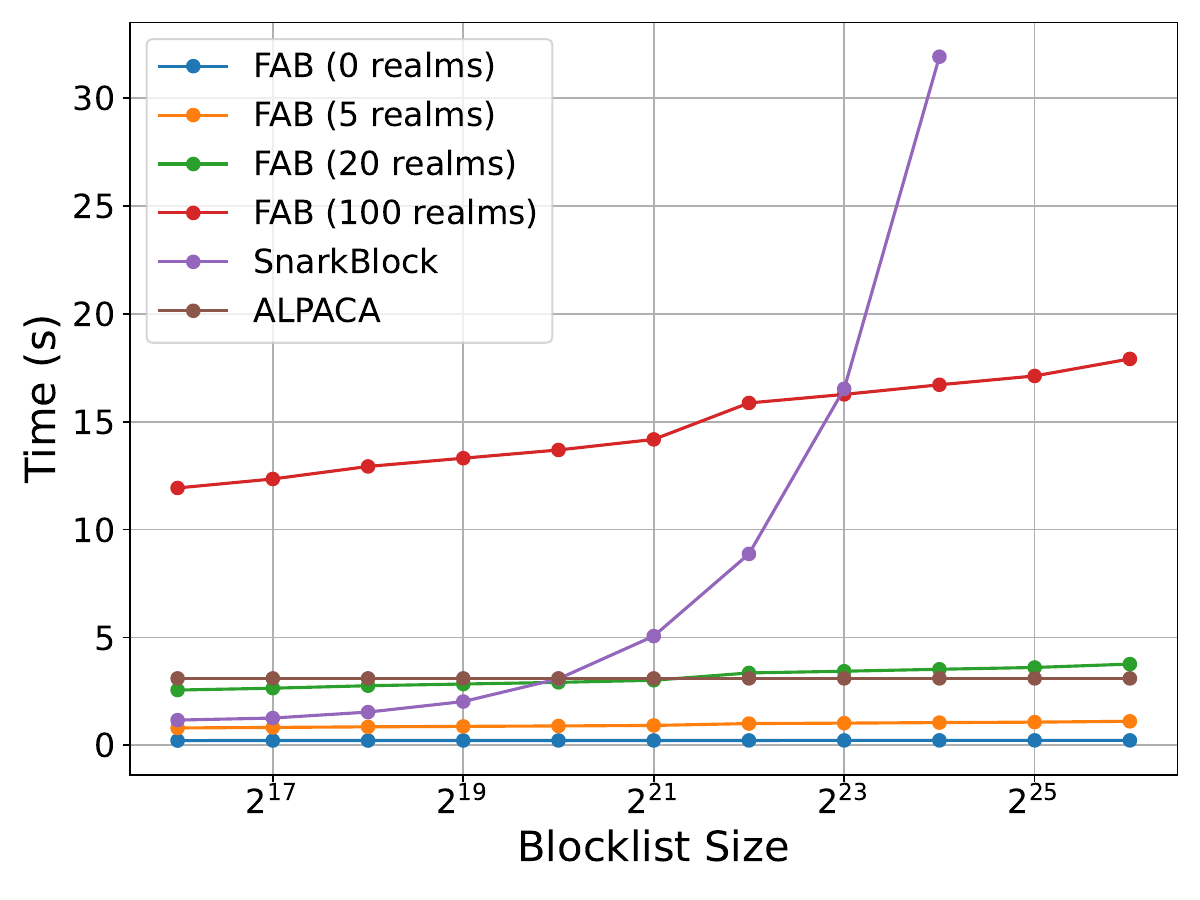}
        \caption{Authentication time.}
        \label{fig:prove}
    \end{subfigure}
    \begin{subfigure}[b]{0.32\textwidth}
        \centering
        \includegraphics[width=1\linewidth]{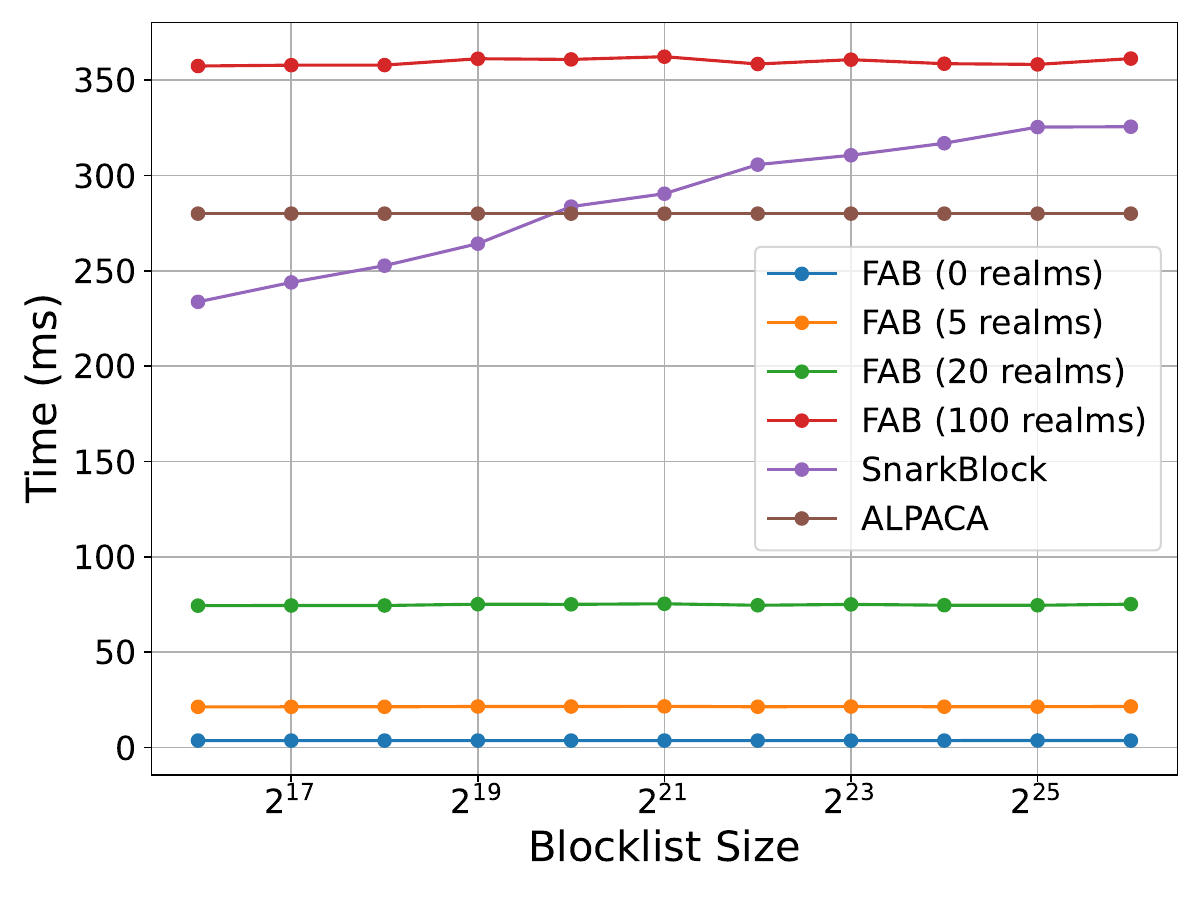}
        \caption{Verification time.}
        \label{fig:verify}
    \end{subfigure}
        \begin{subfigure}[b]{0.32\textwidth}
        \centering
        \includegraphics[width=1\linewidth]{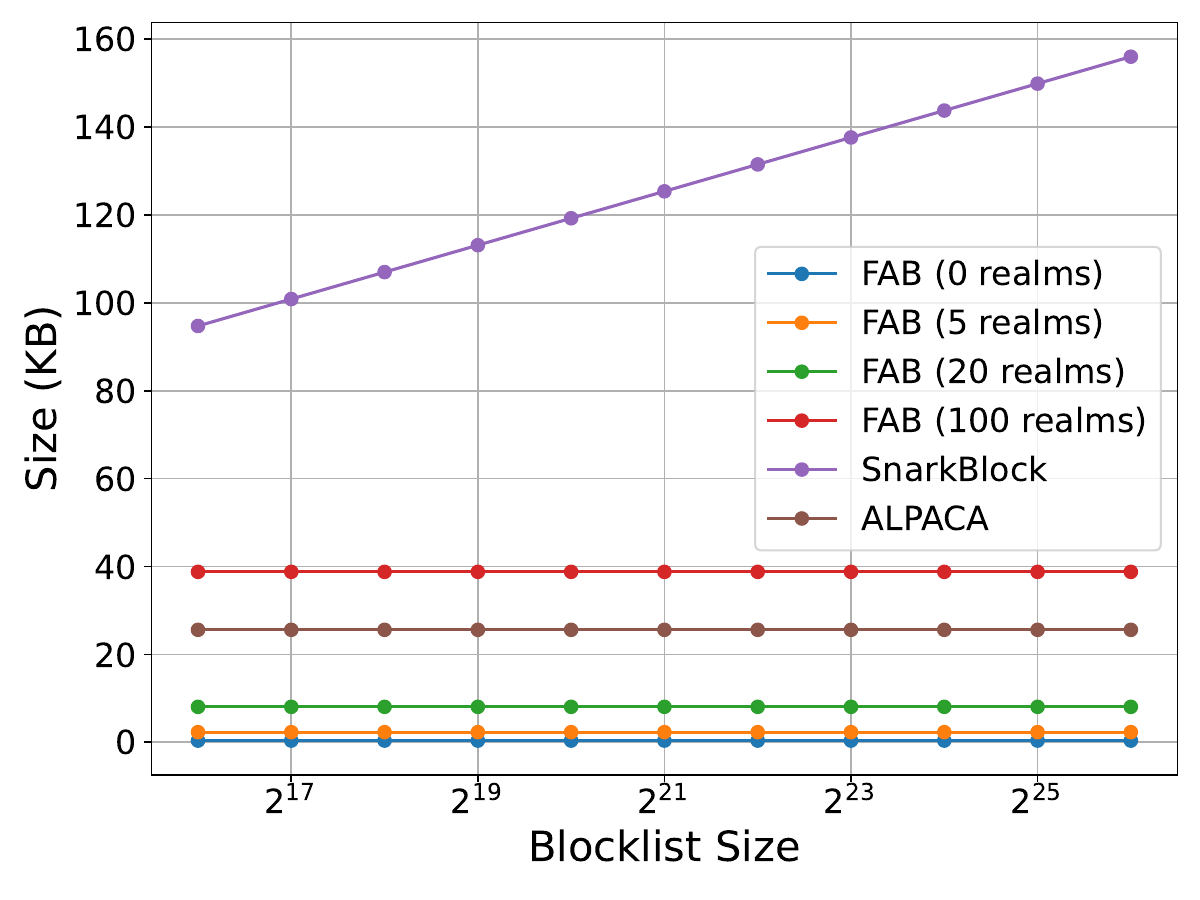}
        \caption{Proof size.}
        \label{fig:size}
    \end{subfigure}
    \caption{Performance evaluation of our FAB scheme and comparison with state-of-the-art Anonymous Blocklisting scheme SNARKBlock and ALPACA. We represent our FAB scheme for multiple amounts of trusted realms. The measurements combine the cost of $R^{auth}$ and $R^{block}$, the latter being executed once for every trusted realm.}
    \label{fig:measurements}
\end{figure*}

Figure~\ref{fig:measurements} shows the performance of our scheme for creating and verifying proofs, as well as their size. The measurements shown include the sum of the costs of $R^{auth}$ and $R^{ban}$. Recall that one $R^{ban}$ must be presented for each trusted realm of the target realm: we represent this by analysing the performance under different amounts of trusted realms, which essentially involves multiplying the cost of $R^{ban}$. The measurements are parametrised by blocklist size; however, we remark that our scheme is not affected by the current size of the blocklist but its \textit{maximum blocklist size}, which needs to be determined as a system parameter. This parameter determines the depth of the Merkle Tree and thus the number of hashes to be executed for $R^{ban}$.


\section{Discussion}
\label{sec:discussion}

We now comment on real-world aspects of our construction.

\subsection{Comparison with other Anonymous Blocklisting schemes}

\begin{table}
\centering
\caption{Comparison of algorithmic complexity between our scheme and SNARKBlock and ALPACA through the parameters that determine the cost of each operation. The performance of offline synchronisation is also accompanied by concrete time measurements. The following abbreviations are used: $BL$: Current blocklist size. $BL_{max}$: Configured maximum blocklist size. $1$: constant operation. $R$: Number of trusted realms of the target realm. \texttt{upd}: Number of updates since last authentication. \texttt{cs}: Chunk size.}
\resizebox{\columnwidth}{!}{
\begin{tabular}{l|lll|l}
\toprule
Work    & Auth & Verify  & Proof Size & Sync (s) \\
\midrule
SNARKBlock & $BL$ & $\log(BL)$ & $\log(BL)$ & $17.87 \lceil \frac{\text{\texttt{upd}}}{ \text{\texttt{cs}}} \rceil$ \\
ALPACA     & $1$ & $1$ & $1$ & $0.39\texttt{upd}$ \\
Ours       & $\log(BL_{max}) R$ & $R$ &  $R$ & N/A \\
\bottomrule
\end{tabular}
}
\label{tab:comparison}  
\end{table}

Table~\ref{tab:comparison} compares the algorithmic complexity of our scheme with the AB schemes SNARKBlock and ALPACA. All of our measurements scale linearly with the number of realms, with Auth times being affected by the depth of the blocklist Merkle Tree. This is an improvement over SNARKBlock, whose performance increases linearly with the blocklist size. Conversely, the complexity of ALPACA is constant: as shown in Figure~\ref{fig:prove}, its authentication cost is roughly equivalent to a FAB authentication with 20 trusted realms.

However, the most significant improvement in our protocol resides on avoiding offline synchronisation costs, also shown in Table~\ref{tab:comparison}. Indeed, processing a single chunk in SNARKBlock is equivalent a FAB authentication with 100 realms. In ALPACA, every new blocked pseudonym must be individually processed with a cost of 0.39 seconds, which becomes the most expensive operation if such updates are common. In contrast, our scheme does not require processing new additions to the blocklist.

\subsection{Advantages and limitations}

Although related, our FAB scheme is based on significantly different assumptions about the structure of service providers than other AB works such as~\cite{blac, snarkblock, alpaca}. These works assume a single service provider with high traffic; indeed, networks with thousands of daily blocks such as Wikipedia or Reddit are often used as examples. 

In contrast, our model considers a high number of comparatively smaller realms. Throughout this document we have used messaging groups to illustrate our envisioned scenario. However, the FAB scheme is also applicable to big Service Providers that are divided into sub-communities, such as Reddit or Discord. Instead of using a single blocklist for the service, each of the communities could represent a different realm in a FAB scheme. This is advantageous as our FAB scheme outperforms other AB schemes in scenarios in which blocks are issued frequently, since it does not require synchronisation. However, the FAB scheme introduces a new linear scaling: the number of trusted realms. If this number is too high, then FAB may become less efficient than traditional AB schemes. Figure \ref{fig:prove} analyses up to 100 trusted realms, resulting in approximately 15 seconds for authentication.

The trust requirements also different for both schemes. In AB, users have no power to insert or remove pseudonyms from the blocklist, and rely on the Service Provider administrators for these activities. In contrast, FAB schemes allow realm creators to dynamically decide whether or not they trust a particular realm, allowing for a more fine-grained definition of trust. This choice on whether or not to trust on other realms could be based on their similarity of rules of conduct, level of moderation or track record. Admittedly, evaluating other realms to decide if they are trustworthy requires an additional effort in group administration. We leave for future work whether or not it is possible to merge or aggregate different realms, which would minimise the cost of such administration tasks.

Lastly, the scheme presented in this work offers Unlinkability across different realms, but not inside the same realm. Indeed, a user must employ the same pseudonym for multiple authentication attempts inside a realm. Traditional AB schemes would be more appropiate in scenarios in which every individual message in a given realm needs to be unlinkable. Our scheme could be modified to provide limited Unlinkability inside a given realm by expanding user credentials to have $n$ identities $x_0, ...x_n$ instead of only one $x$. The modification would allow users to have $n$ possible pseudonyms in a given realm. Unfortunately, this would significantly affect the performance of the $R^{block}$ relation, as it would involve showing that none of the user's $n$ pseudonyms has been blocked, increasing the cost by a factor of $n$. The cost increase could be mitigated by the inclusion of batch proofs~\cite{reckle}, which would allow users to efficiently prove non-membership of their $n$ identities.

\section{Conclusion}
\label{sec:conclusion}

In this work we have presented a novel Federated Anonymous Blocklisting family of protocols that allows unlinkable authentication across Service Providers while introducing the capability of blocking misbehaving users. Unlike similar Anonymous Blocklisting schemes, the performance of our proposed construction does not depend on the size of the blocklist nor requires clients to  process new additions.

Our scheme is particularly useful for environments composed of a large number of independent groups of users that dynamically establish trust relations between each other, such as messaging groups of federated sub-communities. For future work, we plan on adapting our solution to provide unlinkability between multiple authentications inside the same Service Provider while maintaining the advantages in efficiency of our proposed scheme.

\section*{Acknowledgements}

This work has been funded by the European Regional Development Fund (ERDF) through the EU Interreg VI-A Spain-Portugal (POCTEP) 2021-2027 Programme, project "Quantum IBER\_IA: Impulso estratégico de las capacidades en tecnologías cuánticas e inteligencia artificial en el espacio ibérico transfronterizo". D.S. acknowledges support from Xunta de Galicia and the European Union (European Social Fund - ESF) scholarship [ED481A-2023-219]. We also acknowledge support from the Xunta de Galicia and the European Union (FEDER Galicia 2021-2027 Program) Ref. ED431B 2024/21, CITIC ED431G 2023/0.

The authors would like to thank Jiwon Kim for her help in the performance comparison with ALPACA.

\bibliographystyle{IEEEtran}
\bibliography{main}

@misc{messaging-stats,
  author = {Statista},
  title = {Most used communication methods worldwide},
  year = 2025,
  howpublished = {https://www.statista.com/statistics/1459143/most-used-communication-ways-worldwide/},
  urldate = {2025-09-01}
}

@article{e2ee,
  title={On end-to-end encryption},
  author={Hale, Britta and Komlo, Chelsea},
  journal={Cryptology ePrint Archive},
  year={2022}
}

@misc{signal-sender,
      title={Technology Preview: Sealed sender for Signal}, 
      author={Signal},
      year={2018},
      url={https://signal.org/blog/sealed-sender/}, 
}

@inproceedings{franking,
  title={Message franking via committing authenticated encryption},
  author={Grubbs, Paul and Lu, Jiahui and Ristenpart, Thomas},
  booktitle={Annual International Cryptology Conference},
  pages={66--97},
  year={2017},
  organization={Springer}
}

@inproceedings {franking-sealed,
author = {Rawane Issa and Nicolas Alhaddad and Mayank Varia},
title = {Hecate: Abuse Reporting in Secure Messengers with Sealed Sender},
booktitle = {31st USENIX Security Symposium (USENIX Security 22)},
year = {2022},
isbn = {978-1-939133-31-1},
address = {Boston, MA},
pages = {2335--2352},
publisher = {USENIX Association},
month = aug
}

@inproceedings{fediverse,
  title={Will admins cope? Decentralized moderation in the fediverse},
  author={Anaobi, Ishaku Hassan and Raman, Aravindh and Castro, Ignacio and Zia, Haris Bin and Ibosiola, Damilola and Tyson, Gareth},
  booktitle={Proceedings of the ACM Web Conference 2023},
  pages={3109--3120},
  year={2023}
}

@INPROCEEDINGS{ab_formal,
  author={Henry, Ryan and Goldberg, Ian},
  booktitle={2011 IEEE Symposium on Security and Privacy}, 
  title={Formalizing Anonymous Blacklisting Systems}, 
  year={2011},
  volume={},
  number={},
  pages={81-95},
  doi={10.1109/SP.2011.13}
}

@article{blac, 
author = {Tsang, Patrick P. and Au, Man Ho and Kapadia, Apu and Smith, Sean W.},
title = {BLAC: Revoking Repeatedly Misbehaving Anonymous Users without Relying on TTPs},
year = {2010},
issue_date = {December 2010},
publisher = {Association for Computing Machinery},
volume = {13},
number = {4},
issn = {1094-9224},
doi = {10.1145/1880022.1880033},
journal = {ACM Trans. Inf. Syst. Secur.},
month = dec,
articleno = {39},
numpages = {33},
}

@inproceedings{perea,
author = {Tsang, Patrick and Au, Man Ho and Kapadia, Apu and Smith, Sean},
year = {2008},
month = {10},
pages = {333-344},
title = {PEREA: Towards practical TTP-free revocation in anonymous authentication},
journal = {Proceedings of the ACM Conference on Computer and Communications Security},
doi = {10.1145/1455770.1455813}
}

@inproceedings{snarkblock,
	title = {{SNARKBlock}: {Federated} {Anonymous} {Blocklisting} from {Hidden} {Common} {Input} {Aggregate} {Proofs}},
	copyright = {https://doi.org/10.15223/policy-009},
	isbn = {978-1-66541-316-9},
	shorttitle = {{SNARKBlock}},
	url = {https://ieeexplore.ieee.org/document/9833656/},
	doi = {10.1109/SP46214.2022.9833656},
	language = {en},
	booktitle = {2022 {IEEE} {Symposium} on {Security} and {Privacy} ({SP})},
	publisher = {IEEE},
	author = {Rosenberg, Michael and Maller, Mary and Miers, Ian},
	month = may,
	year = {2022},
	pages = {948--965},
}

@article{promises,
	title = {zk-promises: {Anonymous} {Moderation}, {Reputation}, and {Blocking} from {Anonymous} {Credentials} with {Callbacks}},
	journal = {Cryptology ePrint Archive},
	year = 2024,
	language = {en},
	author = {Shih, Maurice and Kailad, Hari and Rosenberg, Michael and Miers, Ian},
}

@article{alpaca,
  title={ALPACA: Anonymous Blocklisting with Constant-Sized Updatable Proofs},
  author={Kim, Jiwon and Kothapalli, Abhiram and Chardouvelis, Orestis and Wahby, Riad S and Grubbs, Paul},
  journal={Cryptology ePrint Archive},
  year={2025}
}

@inproceedings{ivc,
  title={Nova: Recursive zero-knowledge arguments from folding schemes},
  author={Kothapalli, Abhiram and Setty, Srinath and Tzialla, Ioanna},
  booktitle={Annual International Cryptology Conference},
  pages={359--388},
  year={2022},
  organization={Springer}
}

@inproceedings{reckle,
	title = {Reckle {Trees}: {Updatable} {Merkle} {Batch} {Proofs} with {Applications}},
	isbn = {9798400706363},
	shorttitle = {Reckle {Trees}},
	doi = {10.1145/3658644.3670354},
	language = {en},
	booktitle = {Proceedings of the 2024 on {ACM} {SIGSAC} {Conference} on {Computer} and {Communications} {Security}},
	publisher = {ACM},
	author = {Papamanthou, Charalampos and Srinivasan, Shravan and Gailly, Nicolas and Hishon-Rezaizadeh, Ismael and Salumets, Andrus and Golemac, Stjepan},
	month = dec,
	year = {2024},
	pages = {1538--1551},
}

@incollection{oblivious_acc,
	title = {Oblivious {Accumulators}},
	volume = {14602},
	isbn = {978-3-031-57721-5 978-3-031-57722-2},
	language = {en},
	booktitle = {Public-{Key} {Cryptography} – {PKC} 2024},
	publisher = {Springer Nature Switzerland},
	author = {Baldimtsi, Foteini and Karantaidou, Ioanna and Raghuraman, Srinivasan},
	year = {2024},
	doi = {10.1007/978-3-031-57722-2\_4},
	pages = {99--131},
}

@inproceedings{acc,
author = {Barthoulot, Ana\"{\i}s and Blazy, Olivier and Canard, S\'{e}bastien},
title = {Cryptographic Accumulators: New Definitions, Enhanced Security, and Delegatable Proofs},
year = {2024},
isbn = {978-3-031-64380-4},
publisher = {Springer-Verlag},
address = {Berlin, Heidelberg},
doi = {10.1007/978-3-031-64381-1\_5},
booktitle = {Progress in Cryptology - AFRICACRYPT 2024: 15th International Conference on Cryptology in Africa, Douala, Cameroon, July 10–12, 2024, Proceedings},
pages = {94–119},
numpages = {26},
location = {Douala, Cameroon}
}

@inproceedings{acc_survey,
	title = {An {Overview} of {Cryptographic} {Accumulators}},
	doi = {10.5220/0010337806610669},
	language = {en},
	booktitle = {Proceedings of the 7th {International} {Conference} on {Information} {Systems} {Security} and {Privacy}},
	author = {Ozcelik, Ilker and Medury, Sai and Broaddus, Justin and Skjellum, Anthony},
	year = {2021},
	pages = {661--669},
}

@article{acc_survey_2,
	title = {A survey of set accumulators for blockchain systems},
	volume = {49},
	issn = {1574-0137},
	doi = {10.1016/j.cosrev.2023.100570},
	journal = {Computer Science Review},
	author = {Loporchio, Matteo and Bernasconi, Anna and Di Francesco Maesa, Damiano and Ricci, Laura},
	month = aug,
	year = {2023},
	pages = {100570},
}

@article{acc_survey_3,
	title = {Cryptographic {Accumulator} and {Its} {Application}: {A} {Survey}},
	volume = {2022},
	issn = {1939-0122},
	shorttitle = {Cryptographic {Accumulator} and {Its} {Application}},
	doi = {10.1155/2022/5429195},
	language = {en},
	number = {1},
	journal = {Security and Communication Networks},
	author = {Ren, Yongjun and Liu, Xinyu and Wu, Qiang and Wang, Ling and Zhang, Weijian},
	year = {2022},
	pages = {5429195},
}

@incollection{acc_universal,
	title = {Universal {Accumulators} with {Efficient} {Nonmembership} {Proofs}},
	volume = {4521},
	copyright = {http://www.springer.com/tdm},
	isbn = {978-3-540-72737-8 978-3-540-72738-5},
	language = {en},
	booktitle = {Applied {Cryptography} and {Network} {Security}},
	publisher = {Springer Berlin Heidelberg},
	author = {Li, Jiangtao and Li, Ninghui and Xue, Rui},
	year = {2007},
	doi = {10.1007/978-3-540-72738-5\_17},
	pages = {253--269},
}

@article{zcash,
  title={Zcash protocol specification},
  author={Hopwood, Daira and Bowe, Sean and Hornby, Taylor and Wilcox, Nathan},
  journal={GitHub: San Francisco, CA, USA},
  volume={4},
  pages={220},
  year={2016}
}

@misc{bison,
	title = {{BISON}: {Blind} {Identification} with {Stateless} {scOped} {pseudoNyms}},
	shorttitle = {{BISON}},
	doi = {10.48550/arXiv.2406.01518},
	language = {en},
	publisher = {arXiv},
	author = {Heher, Jakob and More, Stefan and Heimberger, Lena},
	month = jul,
	year = {2024},
}

@article{attr_pseudo,
	title = {Attribute {Based} {Pseudonyms}: {Anonymous} and {Linkable} {Scoped} {Credentials}},
	volume = {10},
	issn = {2227-7390},
	shorttitle = {Attribute {Based} {Pseudonyms}},
	doi = {10.3390/math10152548},
	language = {en},
	number = {15},
	journal = {Mathematics},
	author = {Garcia-Grau, Francesc and Herrera-Joancomartí, Jordi and Dorca Josa, Aleix},
	month = jan,
	year = {2022},
	pages = {2548},
}

@inproceedings{signal,
  title={The double ratchet: security notions, proofs, and modularization for the signal protocol},
  author={Alwen, Jo{\"e}l and Coretti, Sandro and Dodis, Yevgeniy},
  booktitle={Annual International Conference on the Theory and Applications of Cryptographic Techniques},
  pages={129--158},
  year={2019},
  organization={Springer}
}

@techreport{mls,
	type = {Request for {Comments}},
	title = {The {Messaging} {Layer} {Security} ({MLS}) {Protocol}},
	url = {https://datatracker.ietf.org/doc/rfc9420},
	number = {RFC 9420},
	urldate = {2024-02-01},
	institution = {Internet Engineering Task Force},
	author = {Barnes, Richard and Beurdouche, Benjamin and Robert, Raphael and Millican, Jon and Omara, Emad and Cohn-Gordon, Katriel},
	month = jul,
	year = {2023},
	doi = {10.17487/RFC9420},
	note = {Num Pages: 132},
}

@techreport{mimi,
	type = {Internet {Draft}},
	title = {More Instant Messaging Interoperability},
	number = {draft-ietf-mimi-protocol-03},
	urldate = {2025-06-19},	
	institution = {Internet Engineering Task Force},
	author = {Barnes, Richard and Hodgson, Matthew and Kohbrok, Konrad and Mahy, Rohan and Ralston, Travis and Robert, Raphael},
	month = mar,
	year = {2025},
}

@techreport{mimimi,
	type = {Internet {Draft}},
	title = {MIMI Metadata Minimalization (MIMIMI)},
	number = {draft-kohbrok-mimi-metadata-minimalization-02},
	urldate = {2025-06-19},	
	institution = {Internet Engineering Task Force},
	author = {Kohbrok, Konrad and Robert, Raphael},
	month = apr,
	year = {2025},
}

@article{aa_cgka,
      title={Attribute-Based Authentication in Secure Group Messaging for Distributed Environments}, 
      author={David Soler and Carlos Dafonte and Manuel Fernández-Veiga and Ana Fernández Vilas and Francisco J. Nóvoa},
      year={2024},
      eprint={2405.12042},
      archivePrefix={arXiv},
      primaryClass={cs.CR},
      journal={arXiv}
}

@misc{orca,
      author = {Nirvan Tyagi and Julia Len and Ian Miers and Thomas Ristenpart},
      title = {Orca: Blocklisting in Sender-Anonymous Messaging},
      howpublished = {Cryptology {ePrint} Archive, Paper 2021/1380},
      year = {2021},
      url = {https://eprint.iacr.org/2021/1380}
}

@inproceedings{cgka_fa,
	address = {Charlotte NC USA},
	title = {Continuous {Group} {Key} {Agreement} with {Flexible} {Authorization} and {Its} {Applications}},
	isbn = {9798400700996},
	doi = {10.1145/3579987.3586570},
	language = {en},
	booktitle = {Proceedings of the 9th {ACM} {International} {Workshop} on {Security} and {Privacy} {Analytics}},
	publisher = {ACM},
	author = {Kajita, Kaisei and Emura, Keita and Ogawa, Kazuto and Nojima, Ryo and Ohtake, Go},
	month = apr,
	year = {2023},
	pages = {3--13},
}

@misc{a_cgka,
      author = {David Balbás and Daniel Collins and Serge Vaudenay},
      title = {Cryptographic Administration for Secure Group Messaging},
      howpublished = {Cryptology ePrint Archive, Paper 2022/1411},
      year = {2022},
}

@inproceedings{poseidon,
  title={Poseidon: A new hash function for $\{$Zero-Knowledge$\}$ proof systems},
  author={Grassi, Lorenzo and Khovratovich, Dmitry and Rechberger, Christian and Roy, Arnab and Schofnegger, Markus},
  booktitle={30th USENIX Security Symposium (USENIX Security 21)},
  pages={519--535},
  year={2021}
}

@InProceedings{groth16,
    author="Groth, Jens",
    editor="Fischlin, Marc
    and Coron, Jean-S{\'e}bastien",
    title="On the Size of Pairing-Based Non-interactive Arguments",
    booktitle="Advances in Cryptology -- EUROCRYPT 2016",
    year="2016",
    publisher="Springer Berlin Heidelberg",
    address="Berlin, Heidelberg",
    pages="305--326",
    isbn="978-3-662-49896-5"
}

@Misc{snarkblock-impl,
  author       = {rozbb},
  howpublished = {\url{https://github.com/rozbb/snarkblock}},
  title        = {SNARKBlock: Federated Anonymous Blocklisting from Hidden Common Input Aggregate Proofs},
  year         = {2021},
  commit       = {3db6736},
  journal      = {GitHub repository},
  publisher    = {GitHub},
}

@Misc{alpaca-impl,
  author       = {jiwonkimpark},
  howpublished = {\url{https://github.com/jiwonkimpark/alpaca}},
  title        = {ALPACA: Anonymous Blocklisting with Constant-Sized Updatable Proofs},
  year         = {2025},
  commit       = {e6515e6},
  journal      = {GitHub repository},
  publisher    = {GitHub},
}

@misc{arkworks,
  author = {arkworks contributors},
  title = {\texttt{arkworks} zkSNARK ecosystem},
  url = {https://arkworks.rs},
  year = {2022},
}

\textbf{David Soler} received his M.S degree in Cybersecurity in University of A Coruña in 2023. His main interests include cybersecurity, applied cryptography, and network communications.

\textbf{Carlos Dafonte} received his Ph.D. in Computer Science from the University of A Coruña in 2002. His research interest includes big data and intelligent systems, computational astronomy and cybersecurity. 

\textbf{Manuel Fernández-Veiga} (SM'10) is with the School of Telecommunications Engineering (University of Vigo, Spain). His research focuses on communication and information theory, with applications in information security, private computation \& learning, and distributed computing. 

\textbf{Ana Fernández Vilas} is Full Professor in the School of Telecommunications Engineering at the University of Vigo. Her current research interest pursue the crystallisation of the theoretical foundation of artificial intelligence in current scenarios and the challenges for quantum networking. 

\textbf{Francsco J. Nóvoa} obtained his Ph.D. degree in computer science at the University of A Coruña, in 2007. His research interests include network security, intrusion detection, QKD, data flow analysis, medical informatics and artificial intelligence.






\end{document}